\documentstyle[multicol,prl,aps,epsf]{revtex}
\begin{document}

\title{Extinction events and species lifetimes in a simple ecological model
}  
\author{Barbara Drossel 
} 
\address{Theoretical Physics Group, University of
Manchester, Manchester M13 9PL, UK} 
\date{\today} 
\maketitle

\begin{abstract}
A model for large-scale evolution recently introduced by Amaral and
Meyer is studied analytically and numerically. Species are located at
different trophic levels and become extinct if their prey becomes
extinct.  It is proved that this model is self-organized critical in
the thermodynamic limit, with an exponent 2 characterizing the size
distribution of extinction events.  The lifetime distribution of
species, cutoffs due to finite-size effects, and other quantities are
evaluated. The relevance of this model to biological evolution is
critically assessed.  
\end{abstract}
\begin{multicols}{2}
In complex systems, not everyone is equal. In a human society,
different individuals fulfil different roles. Similarly, companies
occupy different niches in an economic system, and species occupy
different niches in an ecosystem. Among the rich variety of possible
structures in these systems, power laws take a prominent place and
characterize the size distribution of cities \cite{zipf49}, incomes
\cite{pareto97}, and ecological extinction events
\cite{raup93}. Several models that lead to such a scaling scaling
behavior have been introduced in the literature (for some recent examples see
\cite{zan97,amaral97,sne93}), however, the picture is far from complete.

While reality is certainly best described by a complicated web of
interactions at all levels, the simplest type of models that
incorporate qualitative differences between individuals (or companies
or species) are hierarchical or layered models, where individuals in a
given layer affect individuals in the neighboring layer. Such a
structure can e.g., be found in ecosystems, where species are at
different levels in a food chain, or in economic systems, where
different types of producers are located at different places in a
production chain.  Recently, Amaral and Meyer \cite{amaral98}
introduced a model for large-scale evolution that contains several
trophic levels. Using computer simulations, they found a power-law
size distribution of extinction events. It is the intention of this
paper to prove that this model is indeed critical in the thermodynamic
limit, and to evaluate some of its properties analytically and
numerically. 

The model is defined as follows: Species can occupy niches in a model
ecosystem with $L$ levels in the food chain, and $N$ niches in each
level. Species from the first level $l=0$ do not depend on other
species for their food, while species on the higher levels $l$ feed
each on $k$ or less species in the level $l-1$. Changes in the system
occur due to two processes: (a) Creation of new species with a rate
$\mu$ in each empty niche. If the new species arises in a level $l>0$,
$k$ prey species are chosen at random from the layer below. A species
never changes its prey after this initial choice.  (b) Extinction: At
rate $p$, species in the first level $l=0$ become extinct. Any species
in layer $l=1$ and subsequently in higher levels, for which all preys
have become extinct, also become extinct immediately. This rule leads
to avalanches of extinction that may extend through several layers and
will be shown below to obey a power-law size distribution.
These rules are slightly different from \cite{amaral98}, however, the
results can expected to be the same. 

The dynamics of the model are characterized by slow driving (speciation),
and by rare and fast avalanches (or extinction events). In this respect,
the model is similar to models for sandpiles \cite{bak87}, forest-fires
\cite{dro92}, and earthquakes \cite{ofc92} that are self-organized critical
\cite{bak87} with a power-law size distribution of relaxation events, if
certain conditions are satisfied.

Let us first discuss the case $k=1$, where each species in layers
$l>0$ feeds on one prey species only. In this case, each species is
connected to exactly one species in layer $l=0$ via a food
chain. Since several species can feed on the same prey, the structure
of the ecosystem looks like a set of trees, each consisting of all
species that are connected to the same bottom species.  If a bottom
species has existed for a long time, the tree connected to it extends
through many layers and consists of a large number of species. If a
bottom species is young, only few species of the lowest layers are
connected to it, and the corresponding ``tree'' is small. When a
bottom species becomes extinct, the whole tree of species
connected to it becomes also extinct. Since each bottom species
becomes extinct with the same rate $p$, the
size distribution of extinction events is identical to the size
distribution of trees. 

After some time, the system can be expected to be in a stationary
state where the speciation and extinction rates balance each other
within each layer, leading to a constant mean species densities. Of
course, for a finite system size $N$, there may exist considerable
fluctuations around the mean density. Let $\rho_l$ denote the species
density in layer $l$. The equation of motion for $\rho_l$ is
\begin{equation}
d\rho_l/dt = \mu(1-\rho_l)-p\rho_l\, , \label{em}
\end{equation}
leading to a stationary density $\rho_l=\mu/(p+\mu)$ in each layer. 
The lifetime distribution of species is an exponential,
\begin{equation}
p_T(T) = p \exp(-pT)\, , \label{lifetime}
\end{equation}
and is the same for each species.  Let $s_l^{(i)}(t)$ denote the number
of species in layer $l$ that are connected to bottom species
$i$. Since each newly created species picks its prey species at random
among the existing species in layer $l-1$, the growth of
$s_l^{(i)}(t)$ is given by
\begin{equation}
ds_l^{(i)}/dt=[\mu (1-\rho_l)/\rho_{l-1}]s_{l-1}^{(i)}=p s_{l-1}^{(i)}\,,
\end{equation}
as long as bottom species $i$ does not become extinct.  The
second identity holds in the stationary state.
The size of a ``tree'', $S^{(i)}=\sum_l s_l^{(i)}$ obeys consequently
$
dS^{(i)} / dt = p S^{(i)}$,
leading to 
\begin{equation}
S^{(i)}(t) =  S^{(i)}(0) \exp(pt)\,, \label{growth}
\end{equation}
where the initial tree size is $S^{(i)}(0) = 1$, since only the bottom
species is present (the age of a tree is measured from the creation of
the bottom species on). 

The size distribution of trees, $P(S)$, is related to the age distribution of trees, $P(t)$, via
\begin{equation}
 P(S) = P(t) dt/dS\, . \label{relationSt}
\end{equation}
Now, $P(t)$ obeys the equation 
$dP(t)/dt = -p P(t)\, ,$
or
$ P(t) \propto \exp(-pt)\, ,$
which can be combined with Eqs. (\ref{relationSt}) and (\ref{growth}) to give
\begin{equation}
P(S) \propto S^{-2}\, .
\end{equation}
Since the size distribution of trees is identical to the size
distribution of extinction events, the exponent characterizing the
extinction events is $\tau = 2$, in excellent agreement with the
numerical findings by Amaral and Meyer \cite{amaral98}.

There are two types of finite-size effects that modify this power law:
First, there are effects due to the finite size of $N$, when the
number of layers $L$ is large. Since the height of a tree is
proportional to its age, the number of trees extending up to layer $l$
decrases as $\exp(-l)$, and the mean number of species within layer $l$
that belong to the same tree is proportional to $\exp(l)$. Thus, all species
in layers $l>\ln N $ belong to the same tree and are simultanously
destroyed. Consequently trees cannot become much higher that $\ln N$
layers, and there is a cutoff to the power law $P(S) \propto S^{-2}$
at $S_{max} \sim N \ln N$. If the number of layers $L$
is of the order $\ln N$, the size of the largest extinction event is
of the order $NL$, which is the total system size. Indeed, events of
this size are reported in \cite{amaral98}, where $N=1000$ and $L=6
\simeq \ln N$. (These authors probably have chosen $k>1$.)

Since real ecological systems have only a few trophic level, it is
important to also study the case $L < \ln N$, where the properties of
the system do not depend on $N$ and finite-size effects are due to the
finite number of layers, $L$. Since the height of trees grows linearly
in time, while their size $S$ grows exponentially, the typical size of
the largest trees is now given by $S_{max} \propto \exp{L}$. On the
other hand, each extinction event destroys on an average the same
number of species in each layer, since the same number of species are
created in each layer. Since each extinction event destroys one
species in the bottom layer, the mean number of species destroyed per
extinction event is $\bar S = L$.  Together with the relation
$$ \bar S = \int_0^{s_{max}} S P(S) dS $$ and the assumption that
$P(S)$ obeys a power law, this leads again to $P(S) \propto S^{-2}$
for $S<S_{max}$.
\begin{figure}
\epsfysize=0.35\columnwidth{{\epsfbox{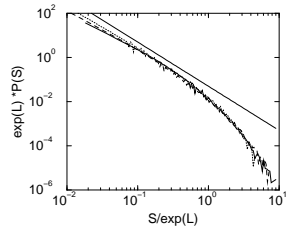}}}
\narrowtext{\caption{The size distribution of extinction events 
for $k=1$, $p = 0.05$, $\mu = 0.02$,  
and $L=4$ (solid line), $L=5$ (dashed line), and $L=6$ (dotted
line). The straight line is a power law with the exponent $\tau = 2$.}}
\label{fig1} 
\end{figure}
Figure 1 shows the size
distribution of extinction events for different values of $L$. The
size $S$ is scaled by $\exp(L)$, and curves for different $L$ collapse
nicely. The number of layers in this simulation is not large enough to
show the power law exponent $\tau$ over a large scaling regime, but it
is chosen such that it is close to the number of layers in real
ecosystems.

The results obtained so far hold for a broad range of values of the
parameters $\mu$ and $p$. Of course, since $\mu$ and $p$ are defined
as rates, one must make sure in simulations with discrete time steps
that they are small enough so that no artificial effects occur due to
the parallel updating of many sites. Also, there is assumed to be a
time scale separation between extinction events which are fast on
evolutionary time scale, and the creation rate of new species, which
is much slower. Only in this case one can neglect the interference
between new speciations and the extinction avalanches.

Let us next discuss the properties of the model in the case
$k>1$. Each newly created species feeds on $k$ species in the layer
below, and becomes extinct only after all $k$ prey species have become
extinct. Older species feed on less prey species and are therefore
more likely to become extinct during a given time interval than
younger species. Also, the prey of older species is in general older
than the prey of younger species, and vanishes faster than the prey of
younger species. In contrast, for $k=1$ each species becomes extinct
with the same probability, irrespective of its age. Choosing $k>1$
introduces correlations in age and extinction probability between
species and their preys that make analytic treatment harder.  

Figure 2 shows the size distibution of extinction events for $k=3$,
$L=6$, and $N=999$. As in a similar plot in \cite{amaral98}, this
size distribution appears to be a power law with extinction events up
to the system size. In order to decide whether the system is indeed in
a critical state, or whether the apparent power law is just due to a
lucky choice of parameters, we have to discuss the system in the
thermodynamic limit $N, L\to \infty$ with $L < \ln N$. 

\begin{figure}
\epsfysize=0.7\columnwidth{{\epsfbox{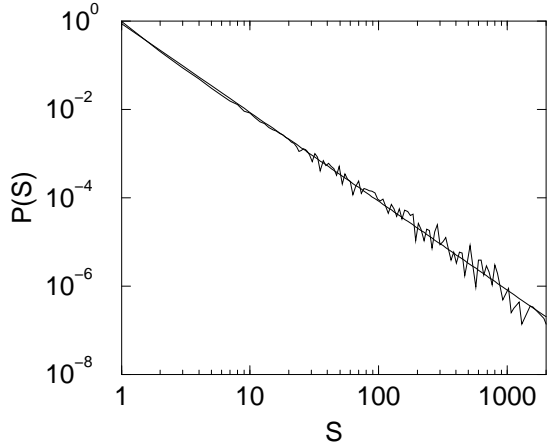}}}
\narrowtext{\caption{The size distribution of extinction events for $k=3$, 
$L=6$, $N=999$, $p=0.05$, and $\mu = 0.01$. The straight line is a power law fit with an exponent $\tau = 2.009$.}}
\label{fig2} 
\end{figure}

Let $P_T^{(l)}(T)$ denote the lifetime distribution of newly created
species in layer $l$. The following calculation relates $P_T^{(l)}(T)$
to $P_T^{(l-1)}(T)$. The preys of a newly created species in layer $l$
are chosen at random from all the species in layer $l-1$. Since
species are created with a constant rate in each layer, the
distribution of the remaining life time $\tau$ of the prey species of
a newly created species is given by $P_{\tau}^{(l-1)}(\tau)$,which is
related to $P_T^{(l-1)}(T)$ via
\begin{eqnarray}
P_{\tau}^{(l-1)}(\tau) &=& \frac{\int_\tau^\infty dT P_T^{(l-1)}(T) }
{\int_0^\infty d\tau \int_\tau^\infty  dT P_T^{(l-1)}(T)}\nonumber \\
&=& \int_\tau^\infty dT P_T^{(l-1)}(T)/ \bar T^{(l-1)}\,, \label{T}
\end{eqnarray}
where $\bar T^{(l-1)}$ is the mean lifetime distribution of species in
layer $l$.  The probability that a species created at time $T=0$ lives
for a time $T$ is identical to the probability that the last of its prey
species becomes extinct at time $T$, leading to
\begin{eqnarray}
P_{T}^{(l)}(T) &=& \left[ 1-\int_T^\infty P_{\tau}^{(l-1)}(\tau) d\tau \right]^{k-1} k P_{\tau}^{(l-1)}(T)\nonumber \\
&=& {\partial \over \partial T}\left[\int_0^T d\tau P_{\tau}^{(l-1)}(\tau) \right]^k.  \label{tau}
\end{eqnarray}
The asymptotic life time distribution of newly created species for
large $l$ is obtained from Eqs. (\ref{T}) and (\ref{tau}) by dropping
the $l$-dependence in $P_T$ and $P_\tau$. After a few steps, one
obtains the following differential equation for $P_\tau$:
\begin{equation}
\frac{dP\tau(\tau)}{[1-\bar T P_\tau(\tau)]^{1-1/k} P\tau(\tau)} = -k d\tau /\bar T.
\end{equation}
Integration of both sides, together with the condition $P_\tau(0) =
1/\bar T$ gives an implicit solution for $P_\tau$. For very large
$\tau$, the first factor in the denominator on the left-hand side
is close to 1 and can be neglected, leading to
$$ P\tau(\tau)\sim \exp(-k \tau/\bar T)$$
for large $\tau$. On the other hand, for very large $\tau$ the life time of a species is only limited by the lifetime of the bottom species to which it is connected. The bottom species is destroyed with a rate $p$, leading to 
$k /\bar T  = p$, or 
\begin{equation}
\bar T = k/p\, . 
\end{equation}
Thus, after a few transient layers, the mean lifetime of newly created
species saturates at $k/p$, which is $k$ times the lifetime of species
in the bottom layer. Figure 3 shows the lifetime distribution of species
in the higher layers for different values of $k$, as obtained by
numerically iterating the recursion relations Eqs. (\ref{T}) and
(\ref{tau}). One can see that the distribution becomes more peaked
with increasing $k$. 

The fact that the life time distribution of species does not change
any more after a few transient layers means that each extinction event
destroys on an average the same number of species on each of the
higher levels. Consequently the rate of species production is also the
same at each of these levels. Let us now define a ``tree'' with index
$m$ to be the set of all species that would go extinct if bottom
species $m$ went extinct. Only species with one prey belong to those
trees, and there are species with one prey that do not belong to a
tree. The growth rate of such a tree must be proportional to its size,
just as in the case $k=1$, since in each layer (above the transitional
layers) species with one prey are generated (from species with two
preys) at the same rate. We can now repeat the derivation of $P(S)$
and $S_{max}$ from above (case $k=1$), and we find again $\tau = 2$
and $S_{max} \sim \exp(L)$. The species density in levels above the
transient levels is $\rho=\mu/(\mu+p/k)$.

\begin{figure}
\epsfysize=0.7\columnwidth{{\epsfbox{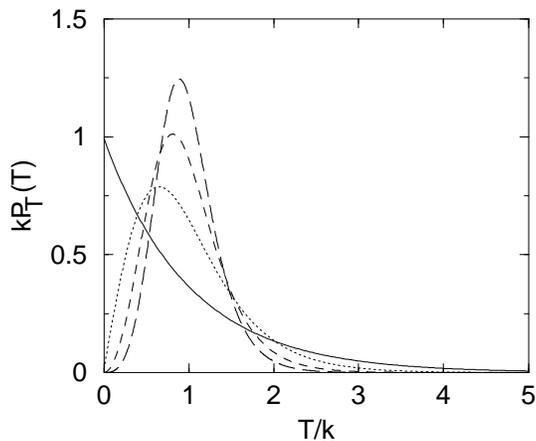}}}
\narrowtext{\caption{The lifetime distribution of species
in the higher layers for $k=1$ (solid), $k=2$ (dotted), $k=3$
(dashed), and $k=4$ (long-dashed). The time axis is scaled by $1/k$.}}
\label{fig3} 
\end{figure}

One of several unrealistic features of this model is that the
number of species does not decrease at higher levels. However, the
biomass must decrease exponentially with the level number, since not
100 percent of the biomass at a given level are consumed by the
species in the next level, and since not all of the consumed prey mass
is turned into predator mass. One might counter this argument by
saying that the dependencies between species are not limited to
predator-prey relationships, and that extinction avalanches will not
only pass from preys to predators, but also to many other
species. This leads, however, to interaction loops instead of nice
hierarchies, and the hierarchical model must be viewed as some crude
mean-field-like approximation to the more complicated reality.

If rule (b) is modified such that a species goes extinct as soon as
one of its $k$ prey species goes extinct, the number of species
decreases exponentially with the level number: Each species is now
connected to $k^l$ bottom species, so that the death rate of species
increases with layer number as $k^l/p$. If the speciation rate is
constant in each layer, the species density decreases as $k^{-l}$,
while the size distribution of extinction events is still a power law
with $\tau=2$ in the thermodynamic limit. However, this power law
cannot be seen for the system sizes used in the simulations. If
the speciation rate is chosen to be proportional to the density of
species in a layer, the density in the stationary state decreases as
$k^{-l(l-1)/2}$, and the size distribution of extinction events has a
cutoff after a few layers, even in the thermodynamic limit.

To summarize, the model discussed in this paper is self-organized
critical with a power-law size distribution of extinction events in
the thermodynamic limit. Finite systems with only a few layers show
this power law only if $k$ is larger than 1 or 2, and a modified
version of the model either does not show a power law in systems with few
layers, or is not critical at all. Thus, power law extinction events
are not a generic feature of food-chain models in general, but occur
only in some versions of these models. Also, a more detailed model
\cite{caldarelli98} that includes adaptation of species to their prey
and that evaluates the transfer of resources from one layer to the
next, was shown to be not critical. 
A model that is completely different from the one discussed in this
paper, but equally simple was introduced some time ago by Bak and
Sneppen \cite{sne93}. This model does not incorporate any layered
structure, but it includes the fitness of species, and it gives a power
law with an exponent different from 2. 

While the study of simple models like the ones mentioned here is a
necessary stage in the attempt to understand complex phenomena like
large-scale evolution, all of them are unrealistic in many respects,
and it can be doubted that they are capable of grasping all
important features of evolutionary dynamics. Certainly, far more
research is needed to gain a deeper understanding of the processes
that lead the the observed patterns in the fossile record.

\acknowledgements 
I thank Alan McKane for interesting discussions. This work was supported by
EPSRC Grant No.~GR/K79307.

\end{multicols} 
\end{document}